\documentclass[epj,nopacs]{svjour}
\usepackage{latexsym}
\usepackage{graphics}
\usepackage{graphicx}
\usepackage{epsfig} 
\usepackage{amsmath}  
\usepackage{paralist}
\pdfoutput=1
\def\mouzz{$^{100}$Mo}
\def\caqo{$^{48}$Ca}
\def\gess{$^{76}$Ge}
\def\seod{$^{82}$Se}
\def\cduus{$^{116}$Cd}
\def\teutz{$^{130}$Te}
\def\xeuts{$^{136}$Xe}

\def\bidud{$^{212}$Bi}
\def\biduq{$^{214}$Bi}
\def\tldzo{$^{208}$Tl}
\def\podud{$^{212}$Po}
\def\poduq{$^{214}$Po}
\def\udto{$^{238}$U}
\def\thdtd{$^{232}$Th}
\def\camo{CaMoO$_4$}
\def\camoenr{$^{40}$Ca$^{100}$MoO$_4$}
\def\camodepl{$^{40}$Ca$^{\mathrm{nat}}$MoO$_4$}
\def\cdwo{CdWO$_4$}
\def\caf{CaF$_2$}
\def\teod{TeO$_2$}

\def\bb{$\beta\beta$}
\def\bbzn{$\beta\beta0\nu$}
\def\bbdn{$\beta\beta2\nu$}
\def\mbb{$|\langle m_{\beta \beta}\rangle |$}
\def\me{$m_e$}
\def\Qbb{$Q_{\beta\beta}$}
\def\FD{${F}_{D}^{0\nu}$}

\def\Tzn{$T_{1/2}^{0\nu}$}
\def\Tdn{$T_{1/2}^{2\nu}$}
\def\rate{counts/(keV$\cdot$kg$\cdot$y)}
\def\vita-dim{$\tau_{1/2}$}

\def\QE{$\epsilon_Q$}
\def\ENCe{$\mathtt{ENC_e}$}
\def\ENC{$\mathtt{ENC}$}
\def\bg{$\beta/\gamma$}

\begin{document}
\title{A flexible scintillation light apparatus for rare event searches}
\author{V. Bonvicini\inst{1} \and S. Capelli\inst{2,3} \and O. Cremonesi\inst{3} \and G. Cucciati\inst{2} \and L. Gironi\inst{2,3} \and M. Pavan\inst{2,3} 
\and E. Previtali\inst{3} \and M. Sisti\inst{2,3}
}                     % Do not remove
%
%\mail{monica.sisti@mib.infn.it}
%
\institute{INFN - Sezione di Trieste, I-34149 Trieste (Italy) \and Dipartimento di Fisica, Universit\`a di Milano-Bicocca, I-20126 Milano (Italy) \and INFN - Sezione di Milano-Bicocca, I-20126 Milano (Italy)}
\date{Received: date / Revised version: date}
\abstract{
Compelling experimental evidences of neutrino oscillations and their implication that neutrinos are massive particles have given neutrinoless double beta decay (\bbzn) a central role in astroparticle physics. In fact, the discovery of this elusive decay would be a major breakthrough, unveiling that neutrino and antineutrino are the same particle and that the lepton number is not conserved. It would also impact our efforts to establish the absolute neutrino mass scale and, ultimately, understand elementary particle interaction unification.
All current experimental programs to search for \bbzn\ are facing with the technical and financial challenge of increasing the experimental mass while maintaining incredibly low levels of spurious background. The new concept described in this paper could be the answer which combines all the features of an ideal experiment: energy resolution, low cost mass scalability, isotope choice flexibility and many powerful handles to make the background negligible. The proposed technology is based on the use of arrays of  %state-of-the-art 
silicon detectors cooled to 120\,K to optimize the collection of the scintillation light emitted by ultra-pure crystals. It is shown that with a 54\,kg array of natural \camo\ scintillation detectors of this type it is possible to yield a competitive sensitivity on the half-life of the \bbzn\ of \mouzz\ as high as $\sim 10^{24}$\,years  in only one year of data taking. The same array made of \camodepl\ scintillation detectors (to get rid of the continuous background coming from the two neutrino double beta decay of \caqo) will instead be capable of achieving the remarkable sensitivity of  $\sim 10^{25}$\,years on the half-life of \mouzz\ \bbzn\ in only one year of measurement. 
\PACS{
      {PACS-key}{discribing text of that key}   \and
      {PACS-key}{discribing text of that key}
     } % end of PACS codes
} %end of abstract
\maketitle
%%%%%%%%%%%%%%%%%%%%%%%%%%%%%%%%%%%%%%%%%%%%%%%%%%%%%%%%%%
\section{Introduction}
\label{intro}
%%%%%%%%%%%%%%%%%%%%%%%%%%%%%%%%%%%%%%%%%%%%%%%%%%%%%%%%%%
In this paper we propose a novel approach to neutrinoless double beta decay searches which uses high performance solid state detectors to read the scintillation light emitted by large mass crystals with high energy resolution. 

Neutrinoless double beta decay is an extremely rare phenomenon hypothesized long time ago but never observed. Its discovery would be a major breakthrough in astroparticle physics, since it would demonstrate the lepton number non-conservation and unveil the Majorana character of the neutrino, i.e. prove that the neutrino is equal to its own anti-particle. At the same time it would allow to assess the absolute neutrino mass scale with high sensitivity and would help point us towards the proper extension of the Standard Model of Particle Physics. This justifies the large number of experimental programs devoted to the search of this decay with different techniques. On the other hand, all these programs are facing with the technical and financial challenge of increasing the experimental mass while maintaining the contribution of the spurious background at incredibly low levels. 

The technique proposed in this paper is the combination in a single device of all the demanding features needed by next generation experiments: high energy resolution, low cost mass scalability, flexibility in the choice of the double beta decaying isotope, and many powerful handles to keep the background negligible. 
Its novelty is the enhancement and optimization of the collection of the scintillation light emitted by ultra-pure crystals through the use of arrays of high performance silicon photodetectors cooled to 120\,K. This would provide scintillation detectors with 1\% level energy resolution and a $\mu$s time resolution. 
We show that with a small array ($\sim 54$\,kg) of such detectors it is possible to yield in only one year of measurement a competitive physics result in the search for the neutrinoless double beta decay of \mouzz\ using natural \camo\ scintillating crystals (i.e. without enriching in \mouzz\ or depleting in \caqo\ isotopes). 
This result would pave the way to a new class of ton scale experiments to search not only for neutrinoless double beta decay with potentially zero background, but also for dark matter and neutrino charged current reactions.

%
%%%%%%%%%%%%%%%%%%%%%%%%%%%%%%%%%%%%%%%%%%%%%%%%%%%%%%%%%%
\section{Neutrinoless double beta decay}
\label{sec:bbzn}
%%%%%%%%%%%%%%%%%%%%%%%%%%%%%%%%%%%%%%%%%%%%%%%%%%%%%%%%%%
In recent years, the discovery of neutrino oscillations unambiguously proved that
neutrinos are massive particles and thus provided the first
confirmation of the existence of new physics beyond the Standard Model \cite{ber12}. 
However oscillation experiments cannot shed light on the fundamental
open issues concerning the absolute mass values or the quantum nature
(Dirac or Majorana fermions) of neutrinos.
Neutrinoless double beta decay is a lepton number violating process with
the unique feature of being the only feasible means to provide insight
into both the above questions. In fact its existence would imply that
neutrinos are massive Majorana fermions and could put important
constraints on the absolute mass scale \cite{ell02,avi08}.

\begin{figure}
\resizebox{0.5\textwidth}{!}{%
  \includegraphics{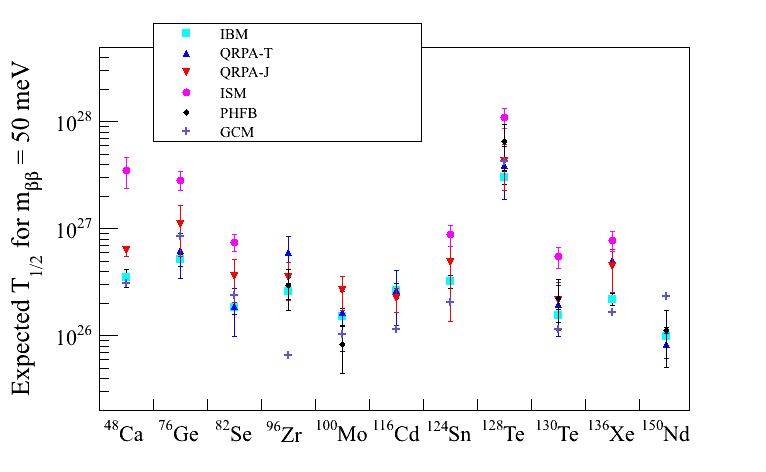}
}
\caption{Neutrinoless double beta decay half-lives
calculated for \mbb=50\,meV
with different most recent theoretical models for several \bbzn\ candidates.}
\label{fig:half-life}       
\end{figure}

Double beta decay (\bb) is a very rare nuclear process in which a
 nucleus (A, Z) decays into its (A, Z+2) isobar. As the expected decay rates 
are extremely low, the choice of the parent nuclei is limited to those which are more bounded than the intermediate ones (to avoid the background caused by the sequence of two single beta decays): this condition is met for a number of even-even nuclei. In the
Standard Model of Particle Physics \bb\ is allowed with the
contemporary emission of 2 electrons and 2 anti-neutrinos (\bbdn),
 and it has indeed been observed experimentally in a
dozen of isotopes with half-lives of the order
$10^{18}-10^{21}$\,y. Non-standard decay
channels appear whenever the Majorana character of the neutrino is
allowed. In this case the lepton number is not conserved and
neutrinoless decay modes are possible. Neutrinoless double beta decay (\bbzn)
can proceed via different mechanisms, the simplest one
being the virtual exchange of a light Majorana neutrino between the
two nucleons. In this case the decay rate is proportional to the square
of the so-called effective Majorana mass \mbb: 
\begin{equation}
\label{eq:half-life}
[T_{1/2}^{0\nu}]^{-1}=\frac{|\langle m_{\beta \beta }\rangle|^{2}}{m_{e}^2}G^{0\nu }|M^{0\nu }|^{2}
\end{equation}
where ${T}_{1/{2}}^{0\nu }$ is the decay half-life, ${G}^{0\nu }$ is the
two-body phase-space integral, ${M}^{0\nu }$ is the \bbzn\
Nuclear Matrix Element (NME), and  \me\
is the electron mass. The product $F_{N}^{0\nu }=G^{0\nu }|M^{0\nu}|^{2}$ includes all the nuclear details of the decay and it is usually
referred to as \textit{nuclear factor of merit}. While ${G}^{0\nu }$
can be calculated with reasonable accuracy, the NME value is strongly
dependent on the nuclear model used for its evaluation. Significant
improvements towards compatibility among various theoretical models
have been achieved in recent years, even if discrepancies of about a
factor 2-3 still exist. Fig.\,\ref{fig:half-life} summarizes the current situation \cite{suh98,sim99,kot12,men09,fae12,fan11,suh12,bar13,rat10,rod10}.

\begin{figure}
\centering
\resizebox{0.35\textwidth}{!}{%
  \includegraphics{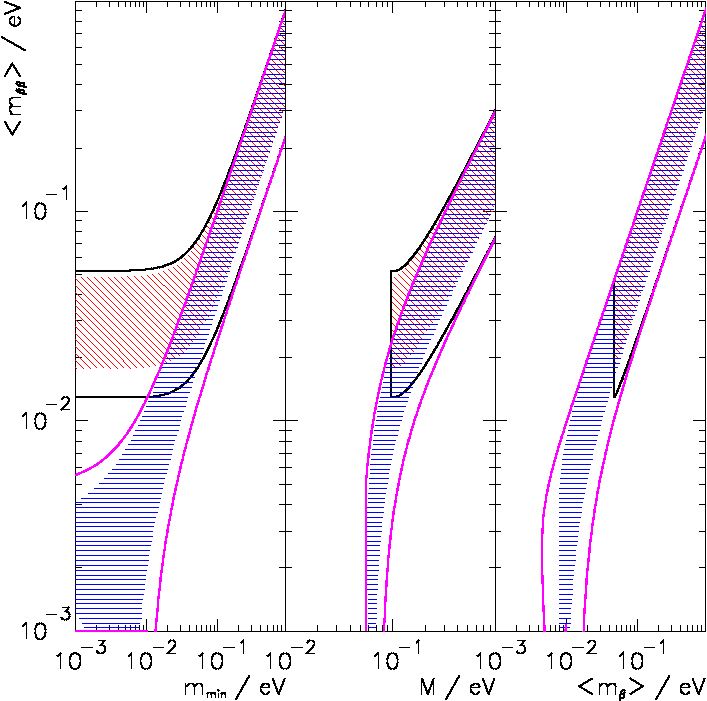}
}
\caption{The left panel shows the expected ranges for \mbb\ as a function of the lightest
neutrino mass. The two sets of solid lines correspond to the Normal (lighter line with horizontal hatch)
and Inverted (darker line with diagonal hatch) Hierarchies, which merge for \mbb $\ge 0.1$\,eV into the degenerate mass pattern. The width of
the hatched areas is due to the unknown Majorana phases and thus
irreducible. The central and right panels show the dependencies of \mbb\ as a
function of the summed neutrino mass and of the kinematical mass 
$|\langle m_{\beta }\rangle |$ (reproduced from \cite{ber12}).}
\label{fig:nu-hierarchies}       
\end{figure}

The effective Majorana mass is a coherent sum over mass eigenstates with
complex Majorana phases and it can be parametrized as a function of neutrino oscillation parameters.
The expected allowed ranges for \mbb\ as
a function of the lightest neutrino mass are depicted in Fig.\,\ref{fig:nu-hierarchies}, where
the solid lines are obtained by taking into account the errors of the
oscillation parameters (at 3${\sigma}$ level). %\cite{ber12}.

Experimentally, \bbzn\ searches rely on the measurement of the two
electron signal. In the so-called homogeneous or calorimetric
experiments the active volume of the detector
contains the double beta decaying isotope (source = detector).
Since the energy of the recoiling nucleus is
negligible, the sum kinetic energy of the two electrons is equal to the
Q-value of the transition (\Qbb). This
monochromatic signal, 
is the main signature used by experiments.
On the other hand, given the rarity of the process, the gathering of the
\bbzn\ counts is hindered by the background events in the energy
region under investigation. The sensitivity of a given \bbzn\
experiment is usually expressed in term of the detector factor of merit \FD,
defined as the process half-life corresponding to
the maximum signal that could be hidden by a background fluctuation
$n_B$ at a given statistical Confidence Level (C.L.). At
1${\sigma}$ level this is given by:
\begin{equation}
\label{eq:FD}
F_{D}^{0\nu }=T_{1/2}^{Back.Fluct}=\ln2 N_{\beta \beta}\frac{T}{n_{B}}=\ln2 \frac{x \eta \epsilon
N_{A}}{A}\sqrt{\frac{MT}{B \Delta}}f(\Delta)
\end{equation}
with  ${n}_{B}=\sqrt{BTM\Delta}$, where \textit{B} is the
background level per unit mass, energy, and time, \textit{M} is the
detector mass, \textit{T} is the measuring time, {${\Delta}$} is
the FWHM energy resolution, $N_{\beta\beta}$ is the number of \bb\
decaying nuclei under observation, \textit{x} is the
stoichiometric multiplicity of  the element containing the \bb\
candidate, ${\eta}$ is the \bb\
candidate isotopic abundance, $N_A$ is
the Avogadro number, \textit{A} is the compound molecular mass, and $\epsilon$
is the detection efficiency. Finally, $f(\Delta)$
is the ``analysis efficiency'', i.e. the fraction of signal
events that fall in an energy window equal to the FWHM ${\Delta}$ centered on \Qbb\
(Region Of Interest or ROI).
The experimental sensitivity \FD\ is then translated into an
effective Majorana mass sensitivity as  \mbb$^{-1}=\sqrt{F_{D}^{0\nu }\cdot F_{N}^{0\nu }}$ showing both the
slow dependence (only with the fourth root) of  \mbb\ on the crucial experimental parameters and how the
isotope choice plays a role in the game from the nuclear model point of
view. 

Apart from one controversial claim of evidence (in tension with several
recent experimental results \cite{exo-200,kamland-zen,gerda}), no \bbzn\ signal has been
detected so far. Experimental half-life lower limits exist for many isotopes. Current best sensitivities have been obtained
for  \gess, \seod, \mouzz, \teutz, and \xeuts, and are
at the level of $10^{24}-10^{25}$\,y, which
translate to  \mbb\ values between 0.2
and 0.7 eV, depending on the \bbzn\ candidate and on the
corresponding NME (see \cite{cre13} for a recent review).

To improve these results several experimental programs are currently on
the way with a variety of challenging techniques and with the common
goal of approaching the Inverted Hierarchy (IH) region of the mass spectrum (see Fig.\,\ref{fig:nu-hierarchies})
in the next years. They require not only very strong efforts towards the
reduction of the radioactive background in the region of interest,
below the intrinsic limit set by the \bbdn\ continuous spectrum, but also good
energy resolutions (approximately better than 2\% at \Qbb \cite{cre13}).
In fact, an approximate expression for the \bbzn\ signal $S$ to \bbdn\ background $B$ ratio in the ROI
can be written as \cite{ell02}: 
\begin{equation}
\label{eq:s/n-bb}
\frac{S}{B}=\frac{m_{e}T_{1/2}^{2\nu }}{7R^{6}Q_{\beta \beta }T_{1/2}^{0\nu}}
\end{equation}
where $R=\Delta /Q_{\beta\beta }$ is the fractional FWHM
resolution. It is then clear that a bad energy resolution can spoil all
background reduction efforts, because of the broadening of the upper tail
of the \bbdn\ continuous spectrum.
Table\,\ref{tab:conteggi-bb} reports the predicted count
rates due to \bbdn\ in a ROI of 50\,keV (therefore assuming a resolution of 1.7\% at 3
MeV) as well as the number of expected \bbzn\ signal events for two different values of \mbb\
(50 and 10\,meV, $3 \sigma$ boundaries of the IH region): the signal $S$ to  background $B$ ratio
 is safely bigger than one for all candidate isotopes. Finally, energy resolution plays
a unique role also when discussing the
discovery potential of any given experimental approach, since it is a crucial parameter to single-out the
\bbzn\ peak over an almost flat background.
\begin{table*}
\centering
\caption{Properties of several \bbzn\ candidates: \textit{Q}-value, isotopic
abundance, \bbdn\ half-life. $R_{2\nu}$ is the rate due to \bbdn\ in a FWHM of 50\,keV centered
around \Qbb\ and is calculated using the Primakoff and Rosen approximation \cite{pri81}. $R_{0\nu}$ is the range of \bbzn\ count rates (from the different NME values of Fig.\,\ref{fig:half-life}) expected for two values of \mbb\ (the effect of the analysis efficiency $f(\Delta)$ has been included). Both $R_{2\nu}$ and $R_{0\nu}$ are expressed in counts per year per ton
of \bbzn\ candidate isotope.}
\label{tab:conteggi-bb}       
\begin{tabular}{ccccccc}
\hline\noalign{\smallskip}
Isotope & \Qbb & $\eta$ & \Tdn & $R_{2\nu}$(50\,keV) & $R_{0\nu}(m_{\beta\beta}=50$\,meV) & $R_{0\nu}(m_{\beta\beta}=10$\,meV)\\
 & [keV] & [\%]   & [y] & [c/y/ton\_iso]       & [c/y/ton\_iso]                      & [c/y/ton\_iso]\\
\noalign{\smallskip}\hline\noalign{\smallskip}
$^{48}$Ca & 4274 & 0.2 & $4.4\times 10^{19}$ & $4.0\times 10^{-2}$ & $1.4-23.1$ & $0.1-0.9$ \\
$^{76}$Ge & 2039 & 7.6 & $1.8\times 10^{21}$ & $2.5\times 10^{-2}$ & $1.2-12.2$ & $0.05-0.5$ \\
$^{82}$Se & 2996 & 8.7 & $9.2\times 10^{19}$ & $6.8\times 10^{-2}$ & $4.4-38.7$ & $0.2-1.5$ \\
$^{96}$Zr & 3348 & 2.8 & $2.3\times 10^{19}$ & $1.3\times 10^{-1}$ & $3.9-50.7$ & $0.2-2.0$ \\
$^{100}$Mo & 3034 & 9.6 & $7.1\times 10^{18}$ & $6.8\times 10^{-1}$ & $8.9-71.3$ & $0.4-2.9$ \\
$^{116}$Cd & 2814 & 7.5 & $2.8\times 10^{19}$ & $2.1\times 10^{-1}$ & $6.8-23.8$ & $0.3-1.0$ \\
$^{130}$Te & 2528 & 34.2 & $6.8\times 10^{20}$ & $1.3\times 10^{-2}$ & $3.6-24.9$ & $0.1-1.0$ \\
$^{136}$Xe & 2458 & 8.9 & $2.1\times 10^{21}$ & $4.8\times 10^{-3}$ & $2.5-14.0$ & $0.1-0.6$ \\
$^{150}$Nd & 3368 & 5.6 & $8.2\times 10^{18}$ & $2.3\times 10^{-1}$ & $9.1-42.7$ & $0.4-1.7$ \\
\noalign{\smallskip}\hline
\end{tabular}
\end{table*}

When the background rate is so low that the expected number of spurious
events in the region of interest is close to zero (\textit{ZB} - Zero
Background condition), the detector factor of merit becomes:
\begin{equation}
\label{eq:FZB}
F_{ZB}^{0\nu }= \ln2 N_{\beta \beta}\frac{T}{n_{L}}=\ln2 \frac{x\eta \epsilon
N_{A}}{A}\frac{MT}{n_{L}}f(\Delta )
\end{equation}
where the background is now described by $n_L$,
 a constant term which represents
the maximum number of counts compatible, at a given confidence level,
with no counts observed \cite{ber12}. In this case the sensitivity to the \bbzn\ signal goes linearly not only with the isotopic fraction and the detection efficiency, but also with the detector mass and the
measuring time, a particularly appealing regime for next generation
experiments.

From the above discussion the strategy for future experimental programs
aiming at exploring the IH region is clear:
\begin{itemize}
\item
minimization of the environmental and cosmogenic radioactive background (experiments
placed underground with heavy shielding, careful selection of
radiopure materials to build the experimental set-up, detector technology with active event identification capability and/or event position reconstruction);
\item
large detector mass, on the tonne scale or larger, and mass scalability;
\item
a well performing detector (good energy resolution and time stability), with flexibility in the choice of the \bbzn\ isotope;
\item
a favourable \bbzn\ candidate isotope (high \Qbb\ value, high natural isotopic
abundance, short half-life, large ratio of \bbzn/\bbdn\ in
the ROI)
\item
a reliable and easy to operate detector technology requiring little service
on long run times;
\end{itemize}
Any of the proposed new projects \cite{cre13} is
a compromise among these often conflicting requests (see also Table\,\ref{tab:cfr-tecniche}).
We show that large arrays of inorganic scintillators could address all the above items in a single approach.

\begin{table*}
\centering
\caption{Main properties of most common calorimetric experimental techniques for \bbzn\ searches.}
\label{tab:cfr-tecniche}       
\begin{tabular}{lcccc}
\hline\noalign{\smallskip}
Experimental approach & FWHM resolution & Mass  & \bbzn\ isotope   & Background reduction \\
 &  $<2\%$ in the ROI & scalability & flexibility &  \\
\noalign{\smallskip}\hline\noalign{\smallskip}
Ge diodes & X &  &  &  $\gamma$ vs. $\alpha/\beta$ (event topology by PSD)\\
Bolometers & X &  & X & $\beta/\gamma$ vs. $\alpha$ (PSD and/or light/heat)\\
Organic scintillators &  & X & X &   \\
Liquid noble gas TPC &  & X & & $\beta/\gamma$ vs. $\alpha$ (light/charge), $\gamma$ vs. $\beta$ (topology)\\
Noble gas TPC & X &  &  & $\beta\beta$ vs. $\alpha/\beta/\gamma$ (event topology)\\
Inorganic scintillators & X & X & X & $\beta/\gamma$ vs. $\alpha$ (PSD), $\gamma$ vs. $\alpha/\beta$ (topology)\\
\noalign{\smallskip}\hline
\end{tabular}
\end{table*}
%
%%%%%%%%%%%%%%%%%%%%%%%%%%%%%%%%%%%%%%%%%%%%%%%%%%%%%%%%%%
\section{The benefit of scintillation detectors}
\label{sec:scintillatori}
%%%%%%%%%%%%%%%%%%%%%%%%%%%%%%%%%%%%%%%%%%%%%%%%%%%%%%%%%%

In realm of \bbzn\ calorimetric experiments, many techniques are available. So far Ge
diode detectors and bolometers have provided excellent energy resolutions \cite{gerda-detect,majorana,cuoricino-finale,cuore-proposal}, at the level of few parts per thousand, and both offer promising solutions for background reduction. On the other
hand, both techniques suffer from limitations in the achievable mass increase
(mainly for cost reasons in the case of Ge diodes, and also technological issues in the
case of bolometers). On the contrary, large scale organic scintillation detectors provide mass
scalability in low-background environment without apparent limitations, but suffer
from a seemingly irreducible restriction in the attainable energy resolution \cite{kamland-zen,sno+}.
Finally, Xe (liquid or gas) based TPC \cite{exo-200,next} are bound to a trade off between source mass and energy resolution,
although they may implement powerful, but complex, background abatement concepts. The possibility to implement in a single approach all the critical features -- high energy resolution, background
abatement techniques and mass scalability -- is the true challenge for next generation projects. Indeed,  
large arrays of inorganic scintillators may be the solution to that.

Inorganic scintillating crystals 
can be grown with high level of intrinsic radiopurity and good
scintillation properties (required for high energy resolution). Moreover, there exist many scintillators containing \bbzn\ candidate nuclei, thus allowing a high flexibility in the choice of the
scintillation detector (i.e. a wide range of \bbzn\ candidates) without the
limitation imposed by other experimental techniques. Furthermore, thanks to Quenching
Factors\footnote{The quenching factor is here defined as the ratio of the
light yield for an alpha particle or a nuclear recoil \ to that of an
electron or a gamma.} (QF) for $\alpha $ particles as small as 0.2 -- which practically remove all
$\alpha $ background from the energy region around \Qbb\ -- and intrinsic $\alpha/\beta$
discrimination ability (Pulse Shape Discrimination or PSD) -- useful to tag $\alpha-\beta$ coincidences -- scintillators
offer powerful tools for background minimization. Finally, large arrays of scintillating crystals can be
assembled with quite simple technological skills and in rather low-cost experimental setups. Despite
all those features, this approach has not exploited its full potential so far, the main reason
being the relatively poor energy resolution achievable with scintillating crystals optically coupled to photomultiplier tubes (see next Section). \\
We discuss in this paper an innovative method to read the scintillation light emitted by large mass scintillators containing the \bbzn\ candidate nuclei without degrading the energy
resolution. In fact, the limit imposed to the scintillator resolution by the carrier statistics alone
is not so crucial. Assuming to have a crystal that emits 10,000 photons per MeV -- a
value definitely reasonable for many scintillators \cite{sci14} -- the relative 
$\mathtt{FWHM}$ resolution $R_{\mathtt{stat}}$ at an energy of 3\,MeV (close to the \Qbb\ of many isotopes) in the hypothesis
of being able to collect all the photons can be estimated as $R_{\mathtt{stat}}(\mathtt{FWHM})=2.355/{\sqrt{3\cdot
10000}}=1.36{\%}$, which corresponds to a $\mathtt{FWHM}$ energy resolution ${\Delta}$ of 41\,keV at 3\,MeV. Even if this value is not comparable to the resolution of Ge diodes or bolometers (which is of the order of 4-6 keV at 3\,MeV), it is not a limiting factor in the case of a low background \bbzn\ experiment (see also Table\,\ref{tab:conteggi-bb}). 
In the following we will demonstrate that it is possible %-- at least in principle -- 
to maintain this resolution performance even with the complete readout chain.
The great advantage of this new method is the availability of a homogeneous detector with the strongest
potentialities towards background reduction and the possibility to
indefinitely scale the experimental mass to boost the sensitivity
towards unexplored regions of the neutrino mass spectrum. This 
detection concept is compared to the other techniques in Table\,\ref{tab:cfr-tecniche}.

%%%%%%%%%%%%%%%%%%%%%%%%%%%%%%%%%%%%%%%%%%%%%%%%%%%%%%%%%%
\section{The measurement technique}
\label{sec:tecnica}
%%%%%%%%%%%%%%%%%%%%%%%%%%%%%%%%%%%%%%%%%%%%%%%%%%%%%%%%%%

The basic idea of the proposed new technique is to optically couple suitable scintillating crystals
to Silicon Drift Detectors (SDD) at a working temperature around 120\,K. Indeed, many
scintillating crystals have been used at low temperatures showing improved light emission
\cite{mik10} with respect to room temperature. On the other hand, in
the recent years there have been many advances in the light detection
using solid state devices \cite{fio13}. The innovative idea is
to combine these features to provide the winning technology for future experiments.\\
Table\,\ref{tab:crist-scint} lists a selection of
scintillating crystals containing interesting \bbzn\ candidate nuclei with transition energy
above the \tldzo\ $\gamma $ peak (the most energetic and intense $\gamma $ line from natural
radioactivity) - crucial issue to fulfill the ZB condition. All of them have fairly well known
scintillation properties at low temperatures \cite{mik10,sci14} and offer
Light Yields (LYs) well above the 10000 ph/MeV value
considered in the previous section. Therefore, all of them are potentially good candidates for high resolution detectors. It is worth stressing here that the development of inorganic  scintillators is a field of continuous innovation that keeps on offering new crystals with competitive characteristics for many applications -- see e.g. \cite{sci14}.

The key point is to avoid a degradation of the resolution (i.e. a loss of carriers) when
the crystal is coupled to a proper device to detect the scintillation
signal. For example, the use of commercial phototubes introduces strong limitations
because of low quantum efficiency \QE\ -- of the order of 40\% in the best cases
-- and rather narrow spectral response, which implies that many of the
crystals of Table\,\ref{tab:crist-scint} do not emit light within the wavelength intervals
required to match the phototube characteristics, thus causing a further
deterioration of the energy resolution.
Moreover it is very difficult to obtain commercial photubes with a low intrinsic radioactivity of the
glass envelope. The use of SDDs as photodetectors represents therefore an appealing possibility,
since these devices are characterized by a \QE\ larger than 80\% in a wide range of
wavelenghts \cite{fio13}.
In addition, silicon is one of the cleanest materials from the radioactivity point of view.\\
To evaluate the energy resolution potentially achievable by coupling SDDs to
scintillating crystals, we must consider the additional contribution of the noise
of the measurement chain:
\begin{equation}
\label{eq:R_tot}
R(\mathtt{FWHM})=\sqrt{R_{\mathtt{stat}}^{2}+R_{\mathtt{noise}}^{2}}=2.355\sqrt{\frac{1}{\epsilon
_{Q}\cdot N_{\mathit{ph}}}+\frac{\mathtt{ENC_e}^{2}}{\epsilon
_{Q}^{2}\cdot N_{\mathit{ph}}^{2}}}
\end{equation}
where \ENCe\ is the equivalent noise charge of the electronic chain and
$N_{ph}$ is total number of photons emitted by the scintillating crystal after
an energy deposition. A competitive detector therefore requires to reduce as much as possible the
term $R_{\mathtt{noise}}$. In practice, one should keep this term at least
within a tenth of the statistical one: 
$\mathtt{ENC_e}=\sqrt{\epsilon _{Q}\cdot
N_{\mathit{ph}}}/{10}=\sqrt{0.8\cdot
30000}/10=16$\,$\mathtt{e}^{-} \mathtt{rms}$, where a LY of 10,000 ph/MeV at an
energy of 3 MeV has been considered. This value of \ENCe\  is
quite demanding but is within reach of presently available technology, as we will discuss in the following Section. 

Another possible approach that can be adopted to readout the signal emitted by the scintillating crystals is the use of a different solid state device: the Silicon Photomultiplier (SiPM). The optimization of this photodetector had a strong development in the last few years. The noise induced by such type of device operated at low temperatures is negligible; on the other hand the main limitations, at the moment, are given by the lower quantum efficiency with respect to the SDDs (due to the larger dead layers that are actually needed to realize a ``large'' surface detector) and by the strong non linearity that can be observed when measuring a large number of photons (due to the intrinsic avalanche signal production in a single cell). Since the optimization of SiPM is expected to continue in the future, it is not excluded that also these devices will became suitable for a rare event search with the same performances that we point out in this paper for the SDDs.
\begin{table}
\centering
\caption{A selection of \bbzn\ candidates contained in
available scintillating crystals. The range of expected half-lives for an
effective mass \mbb\ of 10 meV is reported.}
\label{tab:crist-scint}       
\begin{tabular}{ccl}
\hline\noalign{\smallskip}
Isotope  & \Tzn(10meV)     & Available   \\
	    & [$10^{27}$\,y]  & scintillating crystals\\
\noalign{\smallskip}\hline\noalign{\smallskip}
\caqo   & $7-100$ & \caf, CaWO$_4$, \camo \\
\mouzz  & $1-9$   & \camo, ZnMoO$_4$ \\
\cduus  & $3-10$  & \cdwo, CdMoO$_4$ \\
\noalign{\smallskip}\hline
\end{tabular}
\end{table}

%%%%%%%%%%%%%%%%%%%%%%%%%%%%%%%%%%%%%%%%%%%%%%%%%%%%%%%%%%%
\subsection{Silicon Drift Detectors as photodetectors}
\label{sec:SDDs}
%%%%%%%%%%%%%%%%%%%%%%%%%%%%%%%%%%%%%%%%%%%%%%%%%%%%%%%%%%%

The keystone for the success of the proposed technique is the achievable noise
performance of the SDD detectors.  Considering a solid state device
with capacity $C_d$ coupled to a charge preamplifier with input capacitance $C_i$
and series noise $\langle e_{w}^2 \rangle$, 
the equivalent noise charge can be written as \cite{gat86}:
\begin{equation}
\label{eq:ENC-generale}
\begin{split}
\mathtt{ENC} & = \left[ \frac{k_{1}\cdot \langle e_{w}^{2}\rangle \cdot (C_{d}+C_{i}+C_{p})^{2}}{\tau} \right. \\  
& + k_{3}A_{1/f}(C_{d}+C_{i}+C_{p})^{2} +2k_{2}qI_{l} \tau \bigg] ^{1/2} 
%] ^{1/2} 
\end{split} 
\end{equation}
where $C_p$ is the parasitic capacitance between the detector and the amplifying circuit,
$I_l$ is the leakage current both at the input of the gate of the FET preamplifier and of the solid state
detector, $q$ is the electron charge ($1.6 \times 10^{-19}$\,C), and ${\tau}$
is the shaping time of the acquired signal.
$k_1$, $k_2$ and $k_3$ are parameters related to
the type of shaping form used in
 the  electronic chain (Gaussian, {triangular, trapezoidal, etc.) -- at this level
all of them can be assumed of unit value, also in view of a possible
application of an optimum filter procedure \cite{gat86} as it is actually done in
many experiments. The three terms in the \ENC\ expression are
the so called \textit{series noise}, \textit{1/f noise}, and \textit{parallel noise}, respectively.
As the above formula shows, the achievement of low levels of
\ENC\ requires a careful design and selection of all components:
the detector, the amplifier front-end, and the coupling between the
two.  In particular:
\begin{itemize}
\item 
The contribution of the 1/f noise can be kept at negligible
levels compared to the other terms through a proper choice of the input
device for the preamplifier (for example, a JFET) \cite{gol65}.
\item 
SDDs  are characterized by a very low capacitance $C_d$ of the electrode collecting the signal charge \cite{gat84}, typically of the order of 0.5--1\,pF/cm$^2$; moreover, the front-end electronics can be integrated directly on the same wafer of the photodetector \cite{lec01}, thus minimizing $C_p$.
\item 
The leakage current is one of the most critical parameters. Its
value is related to the size of the solid state detector and is
approximatly proportional to the area of the detector. Typical values
at room temperature are of the order of 1\,nA/cm$^2$ for many silicon devices. Since a desirable order
of magnitude for the detector surface -- to collect the light from a
scintillating crystal devoted to \bbzn\ search -- is of tens of square centimeters, this would result
in a leakage current of several nA that would spoil the \ENC.
On the other hand, measurements on SDDs \cite{fio13} show a marked decrease in the
leakage current with decreasing temperature.
In principle this decrease should have an exponential trend: therefore the leakage current could
become negligible by reducing the operating temperature, although carrier freeze-out sets a lower
limit around 77\,K \cite{pir90}. An extrapolation below $-25\,^\circ$C of the
experimentally observed behaviour of $I_l$ in SDDs shows that a working temperature around 120\,K should be enough to have a leakage current as low as $10^{-14}$\,A per cm$^2$, 
value observed for many semiconductor devices at this temperature.
Moreover, at low temperatures also the leakage
current of the input JFET of the preamplifier becomes negligible: therefore it is advantageous to place
this component at the cold stage of the experimental set-up too.
Another advantage of placing both
the detector and the preamplifier in close vicinity %at low temperatures 
is that the parasitic capacitance $C_p$ can be reduced to small values
($C_p \sim 0.5$\,pF) therefore diminishing its contribution to the first two terms of \ENC.
\item 
The series noise of the preamplifier is given by: $\langle e_{w}^2 \rangle= 2 k_B T \alpha / g_m$
where $k_{B}$ is the Boltzmann constant, \textit{T} is the
operating temperature, $g_m$ is the transconductance of the input JFET, and ${\alpha}$=0.7 for an
ideal JFET. The value of the transconductance increases with the JFET
area, therefore by increasing the size of the device it is possible to
achieve a sensible reduction of $\langle e_{w}^2 \rangle$.
On the other hand this would cause an increase of both the leakage current and of the input
capacitance $C_i$. Also in this case a low operating temperature \cite{gat86}
helps the optimization of the device geometry keeping a
small area to reduce $I_l$ and $C_i$ while controlling $\langle e_{w}^2 \rangle$ with a proper choice of the working temperature, thus achieving a reduction of the total \ENC.
Typical values of $\langle e_{w}^2 \rangle$ at temperatures of about
120-150\,K are of the order of $10^{-18}$\,V$^2$/Hz while for the input
capacitance we can assume $C_i \sim 1$\,pF.
\end{itemize}
In conclusion, an SDD of 1\,cm$^2$ operated at a temperature of about
120-150\,K and coupled to a JFET closely placed are a viable solution for reading out
the light emitted by a scintillating crystal with high
resolution. To evaluate the expected total noise \ENCe\ associated to such
an electronic chain we can  choose a shaping time of about 50 ${\mu}$s, an acceptable value for the
typical decay times of inorganic scintillators, especially at low
temperatures \cite{mik10}. With this selection  of ${\tau}$, together with the values of the
capacitances, $\langle e_{w}^2 \rangle$ and $ I_l$ just inferred, the contributions of the series and of the parallel noises to the total \ENCe\ are comparable:
\begin{equation}
\label{eq:ENCe-singolo}
\begin{split}
\mathtt{ENC_e} &=\frac{\mathtt{ENC}}{1.6\times
10^{-19}} =\frac{\sqrt{8\times 10^{-38}+1.6\times 10^{-37}}}{1.6\times 10^{-19}}\\
&=3.1\, \mathtt{e^{-}} \mathtt{rms}
\end{split}
\end{equation}
This level of noise is for one solid state photodetector of 1\,cm$^2$ of total area.
On the other hand, for applications which need to cover
large areas with light sensitive detectors -- like the one proposed in this paper
-- bigger size SDDs would be necessary. A possible solution -- which may have additional advantages for background reduction, as discussed later -- is to use several independent small area
SDDs to cover completely the surface of interest. The use of
SDDs for segmented readout has already given interesting results for
$\gamma $ and X-ray detection \cite{rac13,che07,fio06}. 
A segmented optical readout has the additional benefit of keeping small drift times within each SDD, i.e. the time needed by the charge generated inside the SDD by an optical photon to reach
the collecting anode. This parameter influences mainly the time response of the SDD. On the other hand, the number of SDDs ($N_{\mathtt{SDD}}$) used to readout the emitted
scintillation light has to be included in the  evaluation of the overall energy resolution: assuming $\sim 40$ SDDs per detector (see later), the additional contribution to the resolution amounts to 
$\sqrt{N_{\mathtt{SDD}}} \times \mathtt{ENC_e}=\sqrt{40}\times 3.1\sim
19.6 \, \mathtt{e^{-}} \mathtt{rms}$, very close the initial goal of $16\, \mathtt{e^{-}} \mathtt{rms}$,
and looks indeed extremely promising.

%%%%%%%%%%%%%%%%%%%%%%%%%%%%%%%%%%%%%%%%%%%%%%%%%%%%%%
\subsection{Scintillating crystal choice}
\label{sec:scelta-crist}
%%%%%%%%%%%%%%%%%%%%%%%%%%%%%%%%%%%%%%%%%%%%%%%%%%%%%%

Among the different scintillating crystals listed in
Table\,\ref{tab:crist-scint}, \camo, \cdwo, and \caf\ look very promising to build large scintillation detector arrays devoted to rare event searches. Their main physical properties are listed in Table\,\ref{tab:proprieta-scintill}. 
In order to give relevance to the discussion, in the following we will refer to \camo\ crystals, even if  it is clear that similar considerations are valid also for \cdwo, \caf, and in general for any scintillating crystal with adequate physical properties.\\
\begin{table*}[t!]
\centering
\caption{Properties of \camo, \cdwo, and \caf(Eu) scintillating crystals. The values at low temperatures are extrapolated from the cited references. The absorption length is an average value of the many different data reported in the literature.}
\label{tab:proprieta-scintill}       
\begin{tabular}{lcccccc}
\hline\noalign{\smallskip}
        & \multicolumn{2}{c}{\camo}   & \multicolumn{2}{c}{\cdwo}  & \multicolumn{2}{c}{\caf(Eu)} \\
Property & 300\,K & 120\,K & 300\,K & 120\,K & 300\,K & 200\,K  \\
\noalign{\smallskip}\hline\noalign{\smallskip}
Atomic mass $\left[ \mathrm{g/mol} \right]$ & 200 & & 360 & &78 & \\
Density $\left[ \mathrm{g/cm^3} \right]$ &$\sim 4.3$ \cite{ann08} &  & 7.9 \cite{bar06}&  & 3.2\,\cite{mik10}& \\
Melting point $\left[ \mathrm{^\circ C} \right]$ &$\sim 1445$ \cite{ann08} & & 1271 \cite{bar06} & & 1418 \cite{saint-gobain}& \\
Lattice structure & Scheelite & & Wolframite & & Fluorite & \\
Energy gap $E_g \left[ \mathrm{eV} \right]$ & 4.0 \cite{mik10} & &4.2 \cite{mik10} &&& \\
Emission maximum $\lambda_{max} \left[ \mathrm{nm} \right]$ &520\,\cite{ann08} & $\sim 530$\,\cite{mik07} &480 \cite{bar06} &480 \cite{mik10} & 435 \cite{saint-gobain} & \\
Light yield $\left[ \mathrm{ph/MeV} \right]$ &$\sim 8900$\,\cite{mik07}& $\sim 25000$\,\cite{mik07} &$\sim 18500$ \cite{mik10} &$\sim 33500$ \cite{mik10} &24000 \cite{kno99}& $\sim 26400$ \cite{ezio-dott}\\
Scintillation decay time $\left[ \mathrm{\mu s} \right]$ &$\sim 18$\,\cite{mik07} &$\sim 190$\,\cite{mik07} & 13 \cite{bar06} &$\sim 22$ \cite{mik10}& 0.9 \cite{kno99}& \\
Refractive index & 1.98\,\cite{mik07}&& 2.2-2.3\,\cite{bar06}&& 1.44\,\cite{kno99}& \\
Absorption length $\left[ \mathrm{cm} \right]$ &$\sim 60$&&$\sim 60$&&& \\
\noalign{\smallskip}\hline
\end{tabular}
\end{table*}

\subsubsection*{\camo}
%%%%%%%%%%%%%%%%%%%%%%%%%%%%%%%%%%%%%%%
\camo\ is an intrinsic scintillator with a high molybdenum content (48\% in mass),
thus extremely attractive for the search of the \bbzn\ of \mouzz, which in
turn  is one of the most interesting isotopes because of its high transition energy and
relatively high natural isotopic abundance (Table\,\ref{tab:conteggi-bb}).
Moreover the relatively low Z of \camo\ reduces its efficiency for $\gamma $ detection, with the largest
$\gamma $ contribution to counts  in the ROI coming from the 3143 keV \biduq\ line (\udto\
decay chain, branching ratio: 0.0012\%). Also the thermal neutron cross section is very low, while the only dangerous cosmogenic
background is the $\beta^{+}$ decay of $^{88}$Y ($Q_{\beta +}=3.62$\,MeV) -- produced by spallation on \mouzz\ -- which anyway has a short half-life ($\tau_{1/2}=106.6$\,d). On the other hand,
among the variety of inorganic materials containing Mo, \camo\ shows a relative
brighter scintillation at room temperature, which increases with lowering the operating
temperature.\\
Concerning the intrinsic radiopurity level attainable in \camo, recent efforts on the careful check
of all raw materials needed for the crystal growing as well as a close control of all
crystallization steps have allowed to reach internal concentrations as low as 10\,$\mu$Bq/kg for both
\udto\ and \thdtd\ \cite{kob13}. Another important feature of \camo\ is a measured QF of 0.2 \cite{ann08}, which prevents the dangerous alpha background from contributing to the ROI.
Moreover, \camo\ offers a pulse-shape discrimination (PSD)
capability between $\gamma $/$\beta$ events and $\alpha$ particles, which is very useful in
the understanding and reduction of background spectra (see later): in \cite{ann08} the demonstrated
discrimination capability was at the level of $\sim$\,90\%, but it was recently measured to be higher than 99.9\% \cite{kob13}.

A drawback in the use of natural \camo\ as detector for the search of the \bbzn\ of \mouzz\
is the presence of the isotope \caqo, which is also a \bbzn\ candidate with a high transition
energy (see Table\,\ref{tab:conteggi-bb}). Despite its very low isotopic abundance, the expected count rate due to the continuous spectrum of the \bbdn\ of \caqo\ in the ROI ($\sim 50$\,keV) of \mouzz\ is at the level of $\sim 10^{-2}$\,\rate. This must be taken into account when conceiving a \bbzn\
experiment aiming at exploring the IH region of the neutrino mass spectrum. In this respect, recent
developments in the production of \camoenr\ crystals, isotopically depleted in \caqo\ and enriched in \mouzz, with scintillation properties very similar to those of the natural crystals \cite{ale11} 
already offer a viable solution to this problem.

%%%%%%%%%%%%%%%%%%%%%%%%%%%%%%%%%%%%%%%%%%%
\section{Expected detector performance}
\label{sec:single-det}
%%%%%%%%%%%%%%%%%%%%%%%%%%%%%%%%%%%%%%%%%%%

A prototype detector single  module to prove this new measuring technique can be conceived as made of a cylindrical scintillating crystal with the two flat circular surfaces optically coupled to two arrays
of SDD photodetectors.
Scintillating crystal indicative dimensions could be 5\,cm of  diameter and 6\,cm of height.  
We imagine each crystal to be laterally covered with a reflective sheet or paint to improve the light collection, and coupled to the SDD array by means of a suitable optical grease.
Given the above crystal dimensions, about 40 exagonal SDDs of $\sim$1\,cm$^2$ of area are needed
to completely cover the two circular surfaces
of the cylinder\footnote{The optimization of the number of photodetectors per crystal surface as
well as of the area of the single SDD must be addressed on the basis of the measured performance
of the solid state detectors.}.

To evaluate the expected energy resolution of a single detector  module, besides the contributions of the
statistical term and of the electronics noise, it is necessary to take
into account also the capability to collect all the generated photons
on the entrance window of the solid state photodetector.
For this purpose, a dedicated Monte Carlo simulation based on the GEANT4 toolkit \cite{ago03} was developed with the main goal of understanding the dependence of the photon collection efficiency on the type of coupling to the scintillating crystal. In the simulation, the scintillator
is described as a cylindrical crystal of \camo\ with the above dimensions and the
characteristics listed in Table\,\ref{tab:proprieta-scintill}.
The lateral crystal surfaces can be chosen to be polished or covered with reflecting materials to vary the probability that a photon is refracted outside the crystal and then
lost. On the two circular surfaces of the cylinder it is possible to
put a coupling layer of 0.1 mm of air (refraction index 1.003) or optical grease
(refraction index 1.6) and, additionaly, a SDD detector with a refraction index of 1.45 and the \QE\ shown in Fig.\,\ref{fig:curva-effic}.
\begin{figure}
\centering
\resizebox{0.4\textwidth}{!}{%
  \includegraphics{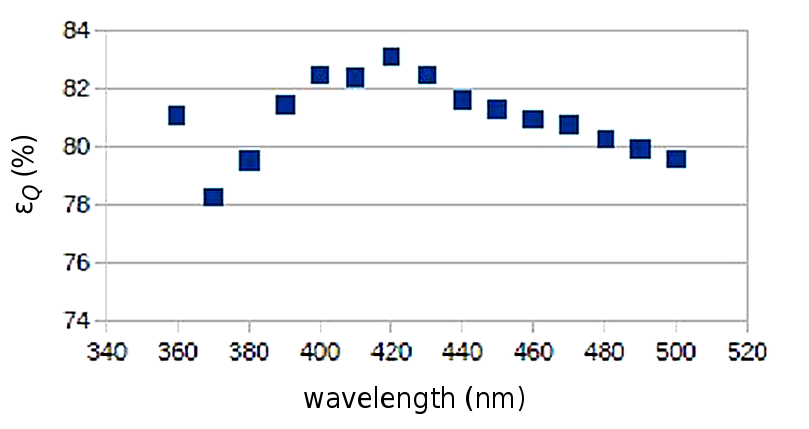}
}
\caption{Quantum efficiency reconstruction from \cite{fio13} for a SDD solid state detector.}
\label{fig:curva-effic}       
\end{figure}
The preliminary simulation results are reported in Table\,\ref{tab:fotoni-ottici}. They show that the fraction of collected light is strongly enhanced by the presence of a grease
interface between the crystal and the silicon device and by the
covering of the lateral crystal surfaces with a diffused
reflector. The fraction of photons generated inside the \camo\ crystal which reach the SDD photodetector
seems very close to 60\% (including the SDD \QE), roughly confirming data available in the
literature \cite{ale13}. These results can be further optimized by using a more precise design of the
photodetectors, of the optical coupling to the scintillator, and of the
crystal lateral surface treatment.
\begin{table*}[t!]
\centering
\caption{Fraction of revealed optical photons depending on the type of crystal lateral surface treatment and on the kind of coupling to the scintillating crystal.}
\label{tab:fotoni-ottici}       
\begin{tabular}{lccc}
\hline\noalign{\smallskip}
\camo  & Coupling layer: &  Coupling layer:   & Coupling layer:   \\
lateral surface treatment  & AIR  & GREASE & GREASE+SDD \\
\noalign{\smallskip}\hline\noalign{\smallskip}
Polished & 0.126 &0.315 & 0.252\\
Polished with specular coating & 0.125 & 0.312 & 0.254\\
Polished with specular wrapping & 0.125 & 0.311 & 0.252 \\
Polished with dffusive coating & 0.361 & 0.692 & 0.562 \\
Polished with dffusive wrapping & 0.126 & 0.312 & 0.255 \\
\noalign{\smallskip}\hline
\end{tabular}
\end{table*}
Finally, the $\mathtt{FWHM}$ resolution attainable at an energy of
3\,MeV with the conceived single detector module operated at 120\,K can be evaluated as:
\begin{equation}
\label{eq:R_tot-finale}
\begin{split}
R(&\mathtt{FWHM})  =\sqrt{R_{\mathtt{int}}^{2}+R_{\mathtt{stat}}^{2}+R_{\mathtt{noise}}^{2}} \\
& \simeq 2.355\sqrt{\frac{1}{\alpha _{\mathit{ph}}N_{\mathit{ph}}\epsilon
_{Q}}+\frac{N_{\mathtt{SDD}}\mathtt{ENC}_{e}^{2}}{\alpha
_{\mathit{ph}}^{2}N_{\mathit{ph}}^{2}\epsilon _{Q}^{2}}}=1.15\text{\%}
\end{split}
\end{equation}
where $\alpha_{ph} \sim$\,70\% is the light collection efficiency just discussed, $N_{ph}\sim$\,75,000
is the total number of photons generated inside a \camo\ crystal operated at 120\,K after an
energy deposition of 3\,MeV, \QE\ $\sim$\,80\% is the SDD quantum efficiency, and 
$\sqrt{N_{\mathtt{SDD}}} \times \mathtt{ENC}_{e} \sim\,20 \,\mathtt{e^{-}} \mathtt{rms}$  is the
equivalent noise charge value calculated above. In the general expression of the
total $\mathtt{FWHM}$ resolution it has been introduced also the contribution
of $R_{\mathtt{int}}$ which is the intrinsic resolution of the scintillator and is
related to several effects, like inhomogeneities due to the local variation of the light output in the
scintillating crystal, variations of the reflectivity of the diffuse reflector
surrounding the scintillator, as well as non-proportionalities of the
scintillator response \cite{dor95,mos05}.
At the energies of interest for \bbzn\ searches ($\sim$\,3\,MeV) the contribution of $R_{\mathtt{int}}$
can be completely neglected, but this assumption must be verified at energies lower than 1\,MeV. 
Therefore the predicted $\mathtt{FWHM}$ resolution at 3\,MeV of a \camo\ detector coupled to 40 SDDs which read the scintillation light is of about 37\,keV: this is a very good value indeed, that allows the minimization of the \mouzz\ \bbdn\ contribution to the ROI as well as a
real discrimination capability of the expected \bbzn\ sharp peak over the background events.

The attainable time resolution of such a detector module is directly connected to the scintillation decay time constant and the total photon emission (Hyman theory \cite{hyman}) through:
$t^{2}=2\tau _{r}\tau_{d}N(t)/N_{\mathit{phe}}$ where $N(t)$ is the total number of
photoelectrons needed to overcome the trigger threshold, $N_{phe}$ is the total number of
photoelectrons collected by the photodetector, $\tau_d$ is the scintillator decay time constant, and
$\tau_r$ is the photodetector rise time. In the case of SDDs, $\tau_r$ is the time needed to drift the
electrons towards the anode ($\sim 1\mu $s). Therefore, using the the data reported in Table\,\ref{tab:proprieta-scintill} for \camo\ at a temperature of 120\,K, and assuming a threshold six time larger than the evaluated $\mathtt{rms}$ noise of 20\,$\mathtt{e^{-}}$,
a time resolution of $\sim 1 \mu$s can be expected (for 1\,MeV energy
deposition), completely dominated by the SDD drift time.

Finally, an additional potentiality of this type of detectors is the possibility of
reconstructing the event topology by exploiting the segmentation of the light readout. In principle the
photodetector matrix placed on the scintillator circular surfaces
should be able to distinguish a single-site event (like the energy
deposition by electrons, as in the case of \bbzn) from a multiple-site event (like an energetic
$\gamma $ interaction) of similar energy. Preliminary Monte Carlo simulations -- which do not include the optical photon propagation yet -- showed that among the $\gamma$  interactions of initial energy equal to 3\,MeV depositing the whole energy in a single \camo\ crystal,
only 30\% would be labelled as single-site event if the threshold distance for distinguishing a multiple-site interaction is set equal to 10 mm; this fraction becomes as low as 14\% if the minimum detectable
inter-site distance is 5 mm. These results are referred to interactions of single $\gamma$s of 3\,MeV: the abatement fraction of the multiple-site events may increase when the 3\,MeV energy is deposited
by coincident $\gamma$s (e.g. coming from the \thdtd\ and \udto\ chains or from showers caused by muons
interacting in the external shielding). These are very preliminary remarks: if confirmed they would open new background discrimination opportunities for this type of detectors.

%%%%%%%%%%%%%%%%%%%%%%%%%%%%%%%%%%%%%%%%%%%%%%%%%%%%%%%%%%%%%%%%%%%%%%%%%%%%%%%%%%
\section{A prototype array of high performance scintillation detectors for \bbzn\ searches}
\label{sec:full-det}
%%%%%%%%%%%%%%%%%%%%%%%%%%%%%%%%%%%%%%%%%%%%%%%%%%%%%%%%%%%%%%%%%%%%%%%%%%%%%%%%%%%

In this Section we investigate the expected performance 
of an array of \camo\ detectors with SDDs' light readout for the search of the \bbzn\ of \mouzz. 
We suppose the full detector to be composed by 108 \camo\ cylindrical
crystals, 5\,cm of diameter and 6\,cm of height, arranged in four layers of 27
single modules for a total mass of 54\,kg. 
We will consider two options:
\begin{inparaenum}[i)]
\item 
an array of 108 natural \camo\ crystals, and
\item 
an array of 108 \camodepl\ crystals, isotopically depleted in \caqo.
\end{inparaenum}
Each scintillator is laterally covered with a reflective
sheet or paint to improve the light collection. The two circular surfaces of every cylindric
crystal are optically coupled to an array of $\sim 20$ SDD light sensors each. The 27
single detectors of any layer are lying side by side and mechanically supported by a
thin copper grid with teflon gaskets. The tentative design is shown in Fig.\,\ref{fig:array}. The full structure can be enclosed in a 5\,cm thick cylindrical copper cryostat of $\sim 45$\,cm of diameter
and $\sim 50$\,cm of height. The Cu vessel is vacuum tight and the operating temperature of 120\,K can be
provided by an external cryocooler thermally coupled to the detector array.
This very simple experimental set-up already reflects many of the shrewdnesses required by a careful background control strategy. In fact, the only
materials in close contact with the detectors are high-purity copper,
teflon, and silicon, besides the wiring. The whole set-up can be
surrounded by at least 20\,cm of lead to shield the experiment
from the environmental radioactivity. 
We imagine this experimental set-up to be located at the Laboratori Nazionali del Gran Sasso, Italy, where an average rock overburden of $\sim 3600$\,m.w.e. shields the experiments from cosmic rays. 
\begin{figure}
\centering
\resizebox{0.4\textwidth}{!}{%
  \includegraphics{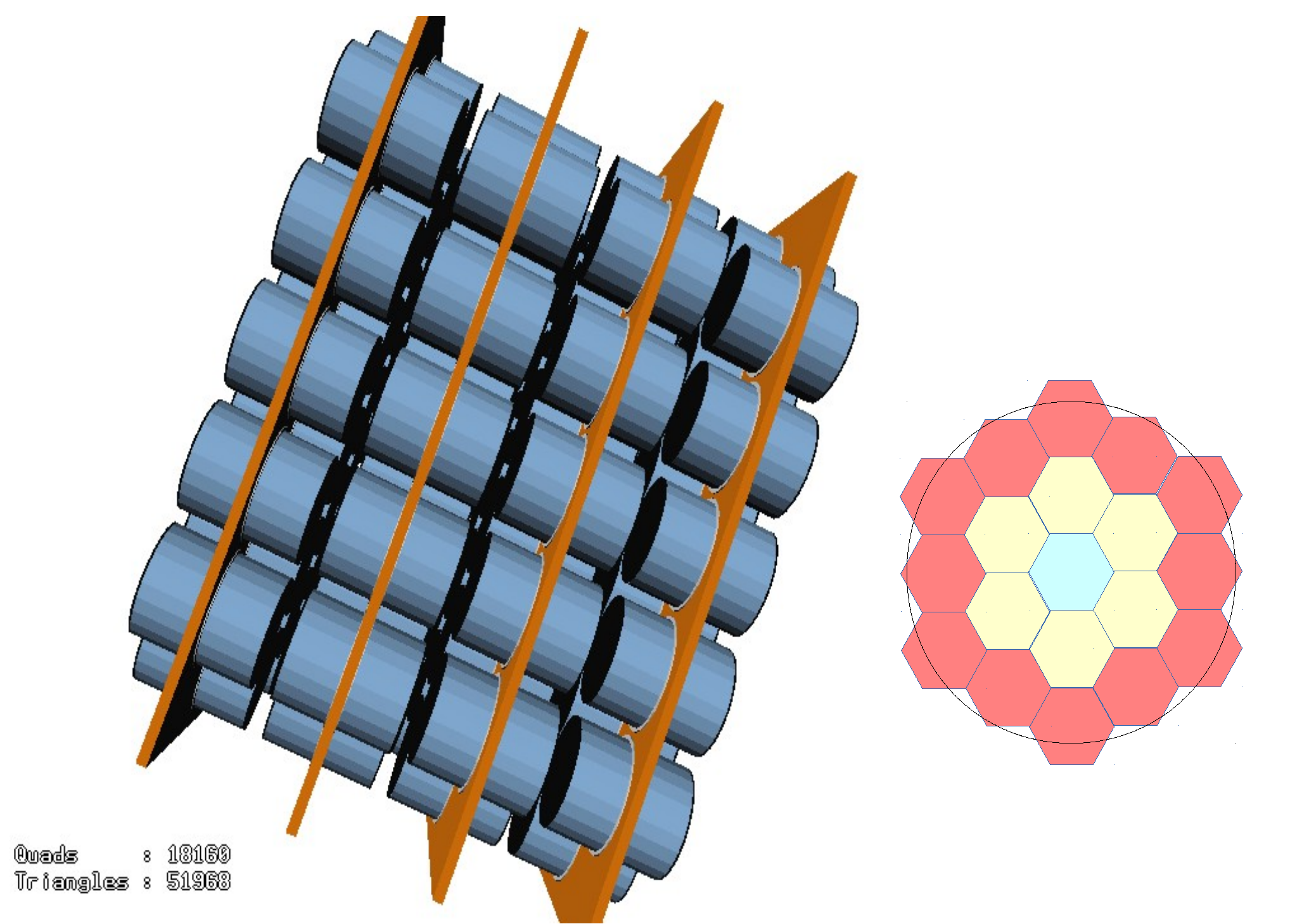}
}
\caption{Array of 108 \camo\ scintillators (left). Possible configuration of the SDD matrix to be otpically coupled to each circular surface of every detector (right).}       
\label{fig:array} 
\end{figure}

In order to reach a competitive sensitivity in a \bbzn\ search, as already
discussed, it is extremely important to remove any spurious source of counts from the ROI.
This is a tricky task, which absorbs most efforts of any experimental programs.
Main sources of background are: 
\begin{inparaenum}[i)]
\item radioactive contaminations of the detector and of the experimental set-up;
\item environmental muons, neutrons and gammas;
\item cosmogenic activation of the detector and set-up materials. 
\end{inparaenum}
The main strategies for background control include a careful selection of the construction
materials, special treatments for reducing surface contaminations of
the elements close to the detectors, adequate shielding against environmental sources, and limited exposure to cosmic rays during construction and transportation.
The above conceived experimental set-up already addresses many of these requests. 
A GEANT4-based Monte Carlo simulation of the whole experimental set-up was performed to
investigate the effect of the possible background sources on the ROI
count rate. In the code, all detector elements can act as a radioactive source with the bulk and surface contaminations independently simulated; for the surfaces,
different contamination depths can be chosen. The following possible sources have been considered in the
simulation: 
\begin{itemize}
\item
\thdtd\ and \udto\ bulk contaminations of the \camo\ crystals and of the Cu (holder\,+\,tank).
Although in the natural radioactive chains there are no intense $\gamma$ lines above 2.615\,MeV, the presence of the two $\beta$ active isotopes \tldzo\ and \biduq\ -- with total decay energies of 4.999 MeV and 3.270 MeV, respectively -- could result in a substantial contribution to the ROI count rate through coincident \bg\ emissions. Moreover, Bi-Po sub-cascades in both \thdtd\ and \udto\ chains -- \bidud\ (\vita-dim =\,60.6\,m) $\beta$ decay to \podud\ (\vita-dim =\,299\,ns), and \biduq\ (\vita-dim =\,19.9\,m) $\beta$ decay to \poduq\ (\vita-dim =\,164.3\,$\mu$s), respectively -- are of even greater concern, since the ``quenched'' $\alpha$ energy from the Po isotopes sums to the energy of the $\beta$ emitted by the parent Bi (because of the very short Po half-life), thus contributing with a continuum
spectrum to the ROI. The assumed bulk contamination limits for copper are available in \cite{art14}. 
For the \camo\ crystals bulk, we have assumed the contamination levels reported in \cite{ann08}. 
This is a very conservative assumption since there are promising purification methods of the raw materials needed for the crystal growing which allow to reduce the intrinsic radioactivity of the scintillators, as already demonstrated in \cite{kob13}. 
\item
\thdtd\ and \udto\ surface contaminations of the \camo\ crystals and of the Cu (holder\,+\,tank).
Surface contaminations very often exceed the background coming from the bulk of the materials, even in
the most radiopure ones: this is supposedly caused by the mechanical
and/or chemical treatments of the materials which may contaminate the surfaces or leave them in a highly  reactive chemical state. The impact of surface contamination on the background depends on the contaminant
and on the depth of the contaminated layer: a depth of $10 \mu$m  has been considered for both the \camo\ crystals and the copper surfaces. The contamination limits for Cu are taken from \cite{art14}, while for \camo\ we have assumed the same limits of \teod\ in \cite{art14}.
\item
\thdtd\ and \udto\ bulk contaminations of the external Pb.
This part of the experimental set-up is placed quite far from the detectors, but it may contribute to the ROI count rate through energetic $\gamma$s from \biduq\ and \tldzo\ (with very low probabilities) or coincident $\gamma$s coming from both natural chains. The probability of such summing becomes negligible as the distance of the emitting source from the detectors increases.
\end{itemize}

The results of the simulation are presented in Table\,\ref{tab:fondo-ROI}: they
have been obtained with the detectors operated in
anticoincidence\,\footnote{This configuration allows to reduce the
background counts by selecting only the events that deposit all their
energy in a single crystal, as it would happen for a \bbzn\
event. Detailed simulations have shown that the
containment efficiency of a \bbzn\ event by a \camo\ crystal
with dimensions similar to those conceived in this paper is $\sim83$\%.}
and by assuming a QF=0.2 for the $\alpha$ particles in \camo.
The time resolution of the detectors has been fixed at 1\,$\mu$s,
the $\mathtt{FWHM}$ resolution at (conservatively) 50\,keV and the threshold for anticoincidence
rejection at 50\,keV.
\begin{table*}[t!]
\centering
\caption{Monte Carlo simulated contribution of the various background sources to the energy
region centered on \mouzz\ \Qbb\ for the full detector array (108 \camo\ detectors). The limits in the last column are obtained by applying \textit{off-line} rejection analyses, as explained in the text.}
\label{tab:fondo-ROI}       
\begin{tabular}{lccc}
\hline\noalign{\smallskip}
Background source & Source & Background in ROI & Background in ROI \\
& activity &&(after \textit{off-line} analysis)   \\
  &  & [\rate] & [\rate] \\
\noalign{\smallskip}\hline\noalign{\smallskip}
\thdtd\ in \camo\ bulk & $<40\, \mu$Bq/kg & $<3\times10^{-1}$ & $<3\times10^{-4}$\\
\udto\ in \camo\ bulk & $<130\, \mu$Bq/kg & $<1\times10^{-2}$ & $<1\times10^{-5}$\\
\thdtd\ on \camo\ surface (10\,$\mu$m deep) & $<2\times10^{-9}$\,Bq/cm$^2$ & $<8\times10^{-6}$ & $<8\times10^{-9}$\\
\udto\ on \camo\ surface (10\,$\mu$m deep) & $< 8 \times 10^{-9}$\,Bq/cm$^2$ & $<1\times10^{-6}$ & $<4\times10^{-7}$\\
\thdtd\ in Cu bulk & $<2\, \mu$Bq/kg & $<1\times10^{-5}$ & \\
\udto\ in Cu bulk & $<65\, \mu$Bq/kg & $<2\times10^{-5}$ & \\
\thdtd\ on Cu surface & $< 3 \times 10^{-8}$\,Bq/cm$^2$ &$<8\times10^{-5}$& \\
\udto\ on Cu surface & $< 3 \times 10^{-8}$\,Bq/cm$^2$ &$<4\times10^{-6}$& \\
\thdtd\ in Pb bulk & $<12\, \mu$Bq/kg & $<9\times10^{-6}$ & \\
\udto\ in Pb bulk & $<74\, \mu$Bq/kg & $<6\times10^{-6}$ & \\
\noalign{\smallskip}\hline
\end{tabular}
\end{table*}

The limits reported for the crystals in the last column of Table\,\ref{tab:fondo-ROI} are obtained with the aid of the $\alpha/\beta$ pulse shape discrimination in \camo\ or by exploiting the technique of delayed coincidences within \thdtd\ and \udto\ radioactive chains. 
In particular:
%
%\begin{inparaenum}[1)]
\begin{enumerate}
\item 
\tldzo\ $\beta$ decay (\vita-dim =\,3.05\,m) induced background can be strongly
reduced by tagging the $\alpha $ particle emitted by the parent nucleus \bidud\ ($\alpha$ decay branching ratio is 36\%): the dead time resulting by vetoing of the detector for ten half-lives of \tldzo\ after every \bidud\ $\alpha$ decay is negligible ($<0.5$\%).
\item 
The background induced by \biduq\ $\beta$ decay (branching ratio\,=\,99.98\%) can be strongly reduced  by tagging the $\alpha$ particle emitted by its daughter nucleus \poduq: the dead time resulting by the removal from the recorded spectrum, of the data segment acquired during the ten \poduq\ half-lives previous to each $\alpha$ decay is completely neglectable.
\item 
The $\alpha + \beta$ pile-up events of the \biduq\ $\rightarrow$ \poduq\ cascades, given the good time resolution of the detectors, can be resolved in 99\% of the cases.
\item
Since the \podud\ half-life is much shorter than the time resolution of the detectors, we have assumed to be able to reject 99\% of the $\alpha + \beta$ pile-up events of the \bidud\ $\rightarrow$ \podud\ cascades thanks to the $\alpha/\beta$ pulse shape discrimination.
%\end{inparaenum}
\end{enumerate}

As Table\,\ref{tab:fondo-ROI} shows, with the highly performing detectors
proposed in this paper the overall background expected in a ROI of 50\,keV centered around the \Qbb\ of \mouzz\ is at the extremely encouraging level of $10^{-4}$\,\rate. \\
This implies that in the hypothesis of an array of 54\,kg of natural \camo\ detectors, the limiting background is the unavoidable \bbdn\ continuous spectrum of \caqo\ (about $10^{-2}$\,\rate). Even in this case, it is thus possible to reach a sensitivity of $\sim 10^{24}$\,y on the half-life of the \bbzn\ of \mouzz\  in only one year of data taking at the Gran Sasso Underground Laboratory -- to be compared with the presently available limit on \mouzz\  ($1.1\times 10^{24}$\,y \cite{nemo3}). \\
In the hypothesis of an array of 54\,kg of \camodepl\ crystals (isotopically depleted in \caqo) operated in the same conditions, the estimated sensitivity on the half-life of  \mouzz\ \bbzn\   after one year of data taking is $\sim 10^{25}$\,y. This shows the tremendous potential of the proposed technique.

%%%%%%%%%%%%%%%%%%%%%%%%%%%%%%%%%%%%%%%%%%%%%%%%%%%%%%%%%%%%%%%%%%%%%%%%%%%%%%%%%%
\section{Other physics measurements with \camo}
\label{sec:other-meas}
%%%%%%%%%%%%%%%%%%%%%%%%%%%%%%%%%%%%%%%%%%%%%%%%%%%%%%%%%%%%%%%%%%%%%%%%%%%%%%%%%%
A \bbzn\ experiment capable of reaching a high enough sensitivity to deeply explore the IH region of the neutrino mass spectrum will require a mass of the candidate isotope of the order of (or higher than) 1\,ton. With an array of \camo\ detectors of this size, interesting topics other than \bbzn\ studies may be at reach.\\
Dark matter may be investigated through the study of the modulation signal. The achievable energy threshold can be evaluated by considering the global light emission of the \camo\ crystals operated at low temperatures coupled to the low electronic noise obtainable with the proposed SDD approach.
As reported in the previous paragraphs, we estimate a global \ENC\ around 20\,$\mathtt{e^{-} rms}$ in normal operating conditions, which means an estimated charge threshold around 100\,$\mathtt{e^{-}}$ considering a Gaussian distribution of the noise fluctuations. Taking into account the photon collection efficiencies reported in Section\,\ref{sec:single-det}, it is possible to estimate an energy threshold of about 6-7\,keV. This threshold value together with the large detector mass and the low background requested by the \bbzn\ experiment,  make the proposed  \camo\ detector suitable for a competitive Dark Matter  experiment. 

Neutrino oscillations and neutrino-nucleus interactions can also be studied through the scattering off Mo nuclei.
Molybdenum isotopes, in fact, are a good target for neutrino interactions. The first suggestion of using a charged current reaction on \mouzz\ to detect solar neutrinos dates back to the year 2000 \cite{ejiri-2000} with a proposal of using a \mouzz\ based detector (MOON experiment) both for \bbzn\ and real-time solar neutrino studies. Recently charged current neutrino reactions on different Mo isotopes have been discussed \cite{suhonen-mo} in view of their use for the detection of supernova neutrinos or for the study of neutrino-nucleus interactions, as suggested in \cite{volpe05}. 
As an example of the potentialities, with the cross sections calculated in \cite{suhonen-mo} for neutrinos of $\sim 1$\,MeV scattering off \mouzz\ nuclei, an interesting interaction rate of $\sim 0.4$\,d$^{-1}$ is expected already on the 54\,kg natural \camo\ crystals array  (discussed in Section\,\ref{sec:full-det}) using the nominal rate for a $^{51}$Cr $\nu_e$ source of 370\,PBq - like the one proposed in \cite{sox} - at 1\,m distance. Obviously this rate will scale linearly with the number of \mouzz\ nuclei, so a proportionally stronger signal is expected by increasing the detector mass and/or by isotopically enriching the \camo\ crystals.

%%%%%%%%%%%%%%%%%%%%%%%%%%%%%%%%%%%%%%%%%%%%%%%%%%%%%%%%%%%%%%%%%%%%%%%%%%%%%%%%%%
\section{Conclusions}
%%%%%%%%%%%%%%%%%%%%%%%%%%%%%%%%%%%%%%%%%%%%%%%%%%%%%%%%%%%%%%%%%%%%%%%%%%%%%%%%%%

In this paper we have discussed a new powerful technology for \bbzn\ searches with the potentialities of achieving zero background in the region of interest.
Adequate radiopure scintillating crystals coupled to low noise SDD photodetectors result in a detector module with all the demanding characteristics required by future \bbzn\ projects: high energy resolution, low cost mass scalability, flexibility in the choice of the isotope, and powerful tools to make the background negligible. 
The proposed concept is therefore ideal for exploring the Inverted Hierarchy region of the neutrino masses, since it allows the measurement of very large detector arrays in quite simple experimental setups. 
The moderate cooling required and the flexibility in the scintillating crystal choice make this approach extremely interesting for many other applications as well. In particular, the \camo\ detector array discussed in this work not only can yield competitive results in the \bbzn\ search of \mouzz\ but can also be used for the study of neutrino oscillations and scattering and of dark matter annual modulation, for a high sensitivity \textit{multipurpose} experiment.

%\bigskip
%\end{document}

\end{document}